# Label-free SERS Discrimination of Proline from Hydroxylated Proline at Single-molecule Level Assisted by a Deep Learning Model

Yingqi Zhao[1,3] Kuo Zhan [1,3], Pei-Lin Xin [1,3], Zuyan Chen[4], Shuai Li[4], Francesco De Angelis[5]  and Jianan Huang [1,2,3,*]

[1] Research Unit of Health Sciences and Technology, Faculty of Medicine, University of Oulu, Aapistie 5 A, 90220 Oulu, Finland.

[2] Research Unit of Disease Networks, Faculty of Biochemistry and Molecular Medicine, University of Oulu, Aapistie 5 A, 90220 Oulu, Finland.

[3] Biocenter Oulu, University of Oulu, Aapistie 5 A, 90220 Oulu, Finland.

4 The Biomimetics and Intelligent Systems (BISG) research unit, Faculty of Information Technology and Electronic Engineering, University of Oulu, Oulu, Finland

5 Istituto Italiano di Tecnologia, Via Morego 30, 16163, Genoa, Italy.

*Email: jianan.huang@oulu.fi

**ABSTRACT:** Discriminating the low-abundance hydroxylated proline from hydroxylated proline is crucial for monitoring diseases and evaluating therapeutic outcomes that require single-molecule sensors. While the plasmonic nanopore sensor can detect the hydroxylation with single-molecule sensitivity by surface enhanced Raman spectroscopy (SERS), it suffers from intrinsic fluctuations of single-molecule signals as well as strong interference from citrates. Here, we used the occurrence frequency histogram of the single-molecule SERS peaks to extract overall dataset spectral features, overcome the signal fluctuations and investigate the citrate-replaced plasmonic nanopore sensors for clean and distinguishable signals of proline and hydroxylated proline. By ligand exchange of the citrates by analyte molecules, the representative peaks of citrates decreased with incubation time, proving occupation of the plasmonic hot spot by the analytes. As a result, the discrimination of the single-molecule SERS signals of proline and hydroxylated proline was possible with the convolutional neural network model with 96.6% accuracy.

## INTRODUCTION

Detecting proteins and their post-translational modification (PTM) is essential to proteomics and diagnosis. PTM plays a vital role in controlling cellular processes by significantly affecting the structure and dynamics of protein. Proline(Pro) and its hydroxylation play a key role in collagen stability. For example, the hydroxylation of two proline residues in hypoxia-inducible factor 1 alpha (HIF-1α) results in proteasome degradation.[1, 2] Analysis of trace amounts of PTM biomarkers in human biofluids is of great significance for early-stage disease diagnosis and low-abundance protein studies.[3] The single-molecule detection of an amino acid and its PTM is the foundation of discriminating PTM sites in peptide and protein PTM analysis.

Currently, the mainstream technology for PTM analysis is mass spectrometry, but its sensitivity is limited by the requirement of $10^6$-$10^8$ copies of molecules. PTMs identification using mass spectrometry also suffers from false identification due to the change of the combinatorial rules for spectra explanation. One emerging single-molecule sensor is the nanopore resistive pulse sensor, which distinguishes biomolecules according to the blockage current changed when the molecules flow through a nanopore. By monitoring the current feature change, single-molecule PTMs discrimination achieved in nanopores, including phosphorylation,[4] acetylation,[5] propionylation,[6] glycosylation,[7] nitration and oxidation.[8] However, other smaller PTMs, such as hydroxylation, can not induce sufficient current feature change in the side nanopore for detection. Therefore, hydroxylation detection at the single molecule level remains a challenging task.

Surface-enhanced Raman spectroscopy (SERS), which provides excellent sensitivity and molecule structure information[9], was believed to be a potentially helpful tool for label-free PTM detection and, therefore, attracted considerable research efforts.[2] Numerous research works focused on larger PTM sites with large Raman cross-sections or distinct bands in spectra, for example, phosphorylation[10, 11], nitration[12], and oxidation.[13] Yet, fewer SERS-based detections were reported for the PTM sites with small Raman cross-sections, such as hydroxylation, due to the difficulties in obtaining sufficient spectra for analysis or the limitation of SERS system sensitivity.[14, 15] Another problem caused by the small PTM Raman cross-section is that the backbone signal may cover or merge the PTM information.[2] Other challenges in SERS PTM detection lie in the interpretation of spectra. Traditionally, molecule discrimination was based on spectra comparison and peak assignment, which provide information about structure differences. Many SERS systems, especially those that depend on the hotspots in nanogaps, usually suffer from unrepeatable spectra, which hinder reliable peak assignment. To date, there has been no report on Raman-based PTM detection at the single-molecule level.

In contrast to the above methods, our recent plasmonic particle-in-pore technology demonstrated Raman spectroscopic detection of the 20 amino acids at the single-molecule level.[16, 17] By constructing a plasmonic nanogap between a gold nanoparticle and nanopore side wall, the particle-in-pore system enabled the shaping of a plasmonic hot spot with single-molecule sensitivity. The hotspot size is comparable with the size of amino acids; when the amino acid molecule enters the hot spot, part of the molecule is located in the hotspot and is excited[18], thereby generating a SERS signal from the corresponding molecule moiety and avoiding signal coverage. The narrow peak width[17] of single-molecule Raman peaks avoided the signal merge as the Raman signals in conventional SERS systems.

In this paper, we report SERS discrimination of proline and hydroxyproline(Hyp) at a single molecule level using particle-in-pore technology with the assistance of a 1-dimensional(1D) convolutional neural network (CNN) model. To the best of our knowledge, this is the first time amino acid with and without hydroxylation, which is one of the most challenging PTM for SERS, is discriminated at a single molecule level. To visualize a large amount of inhomogeneous and vibrating single-molecule data, Raman peak occurring frequency histograms were employed to filter peak intensity, which was mainly influenced by nanostructure morphology and monitor the peak position distribution, which provides structural information for molecule discrimination. The process of analyte substituting citrate on the gold nanoparticles was also investigated with the aid of frequency histograms to ensure a sufficient analyte substitution and minimized citrate(Cit) interference on the spectral interpretation and following deep learning model training. The 1D CNN model was trained to assist with spectra discrimination and demonstrated accuracy above 96% in both testing and post-evaluation. We also performed feature extraction by 1D gradient-weighted feature visualization to localize Raman bands corresponding to spectral change induced by the hydroxy group on the pyridine ring.

Experimental Section

Materials

Nonfunctionalized gold nanoparticles (AuNPs) with an average particle size of 50 nm from Sigma( 753645-25ML, concentration of $3.5 \times 10^{10}$ particles/mL ). SYLGARD™ 184 Silicone Elastomer Kit was used for Polydimethylsiloxane (PDMS) microfluidic channel fabrication. Si wafers with 100nm SiN membranes coated on the surface were purchased from MicroChemicals GmbH. Trans-4-Hydroxy-L-proline(H54409) and L-Proline were purchased from Sigma.

Nanopore Device Fabrication

The gold nanoholes were fabricated on low-stress SiN membranes supported on silicon. The size and thickness of the SiN window were 1 × 1mm and 100nm, respectively. The size of silicon chips which support the SiN window is 1cm ×1 cm. After sputtering a 2 nm titanium and 100 nm gold layer on the front side and 20nm gold layer on the back side of the SiN membrane, focused ion beam(FIB) milling (FEI Helios DualBeam) from the back side of the membrane to create nanopores with 200nm diameter. A scanning electron microscope(SEM) was used to characterise nanopore size and morphology from the front side. Then, the nanopore samples were embedded in a homemade microfluidic chamber made from PDMS.

Attachment of amino acid on AuNPs

All the amino acids used in the measurements were attached physically to gold nanoparticles, following the protocol in our previous work.[16, 17] In the final solution for Raman measurements, the concentration of AuNPs and the salt concentration were $1.3 \times 10^{10}$ particles per mL and 5% of pH 5.5 PBS buffer. Amino acid or peptide stock solutions were mixed with gold nanoparticles and PBS buffer. Before Raman measurement, the mixture was kept in a refrigerator at 4 °C for 48h to allow the adsorption of analytes on AuNP. The concentrations of amino acids and peptides in the final solution were calculated according to previous literature[19] To ensure the number of molecules adsorbed on each AuNP forms a monolayer, in the particle-in-pore system, only one molecule occupies the hot spot and generates a single-molecule SERS signal. The details of the concentration calculation are in Supporting Information S1.

Raman measurement

Raman measurement was performed using a Thermo Scientific DXR2xi Raman Imaging Microscope under a 785nm laser with a 60x water immersion objective. The spectra accumulation time was 0.1s, and the laser power was 10mW. The time series Raman spectra were collected by Andor Solis. Before measurement, the nanopore devices were treated with O2 plasma to generate a hydrophilic surface and pore side walls. Then AuNP was loaded with the analyte, and 5% PBS with pH 5.5 was filled in the bottom and top chamber of the nanopore device. Proline and hydroxyproline were tested in different devices. A new sensor device was used for each AuNP incubation time. During the measurement, 0-0.6V cross SiN membrane voltage was applied using platinum electrodes to facilitate the gold nanoparticle trapping in the nanopore. Signals were collected from more than five different nanopores on the same device.

Raman Data processing

Raman data analysis was performed using MATLAB R2022a software. All the raw data were pre-processed by performing cosmic ray removal, normalization, and baseline reduction. Peak findings were then carried out to locate the peak position using the "find peak" function in MATLAB. A minimum peak intensity threshold of 0.07 was set to select the peak from the noise. The spectra with found peaks were regarded as effective and recorded for further analysis. The number of raw spectra and effective spectra are listed in the form S2 in the supporting information. The normalized peak occurring frequency was calculated by counting the peak occurring at each Raman shift and dividing by the number of effective spectra.

CNN Model

We used MATLAB to run the codes of the CNN model for classification and post-evaluation. A total of 19000 and 500 peak assignment selected spectra with 1463 features were classified using 5-fold cross-validation for the CNN classification model and post-evaluation model, where 80% of the spectra were used as the training set and 20% of the spectra were used as the test set, following these steps: 1)Dataset Preparation, 2) Training Process, 3)Evaluation Metrics. The detailed codes are seen in the supporting information.

1D gradient-weighted feature visualization was adopted to extract outputs normalized feature gradients, providing insights into the role of specific features in model predictions. The approach utilizes feature gradients to quantify the contribution of individual features to a trained CNN model, facilitating interpretability and model diagnostics. Following these steps: 1. Pre-Trained Network Modification The trained CNN model was loaded, and its architecture was modified to allow gradient computation. 2. Data Preparation Features for the first sample (1:1463) were extracted and normalized using pre-computed normalization parameters 3. Gradient computation. Gradients of the loss concerning input features were calculated for each sample. The loss function was defined as the categorical cross-entropy between predicted and true labels. Gradients were computed using a custom function. The detailed codes are seen in the supporting information.

Result and Discussion

Figure 1(a) shows the schematic illustration of the particle-in-pore system. It consists of a gold nanohole array with a 200nm diameter drilled on a 100nm gold film supported on a free-standing SiN membrane with a silicon frame. The SEM image of a typical nanopore is shown in Figure 1 (b). The SiN membrane chip is embedded in a microfluidic device with a top and bottom reservoir. The analyte was physically adsorbed on 50nm diameter gold nanoparticles(as shown in Figure 1(c)) and injected in the bottom reservoir of the microfluidic chamber. Then, the negatively charged AuNP were driven by a cross-membrane potential from the bottom chamber through the nanopore. The nanopore is illuminated from the top with a 785nm laser, which generates optical force and pushes the nanoparticle towards the nanopore sidewall. With the combination of electrophoresis force, electroosmosis force and optical force, the particle will be pressed and trapped near the sidewall for a prolonged time.[16] The analyte enters the hot spot formed between the AuNP and nanopore sidewall, will be excited by an extremely strong localized electromagnetic field, and generate a Raman signal.

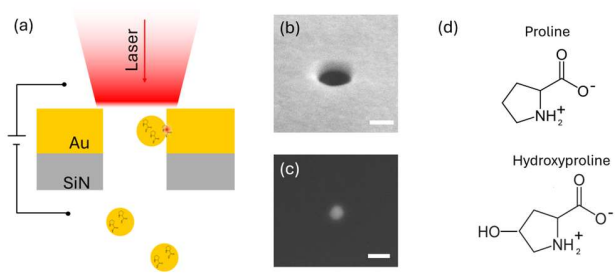

*Figure 1. (a) Schematic illustration of the particle-in-pore sensor that allows single AuNPs loaded with analytes to be trapped in a gold nanopore with plasmonic resonance upon laser excitation at 785 nm. (b)SEM image of the nanopore. The scale bar is 200nm. (c)SEM image of the AuNP. The scale bar is 100nm. (d)The molecule structure of proline and hydroxyproline at pH5.5 in solution.*

Proline and hydroxyproline monolayer were attached to AuNP in 5% PBS buffer with pH 5.5. Both molecules are in their zwitterion form at this pH, as shown in Figure 1(d). The details of the analysis of the charging status are discussed in supporting information S1. The purchased AuNPs have citrate on the surface and serve as a stabilizing surfactant. By incubating the AuNP in proline or hydroxyproline solution, the citrate will be substituted by amino acids since it has lower binding affinity compared to amino acids.[20],[21] The concentration of the incubating solution was calculated according to the AuNP surface area and molecule area to form a monolayer capping.

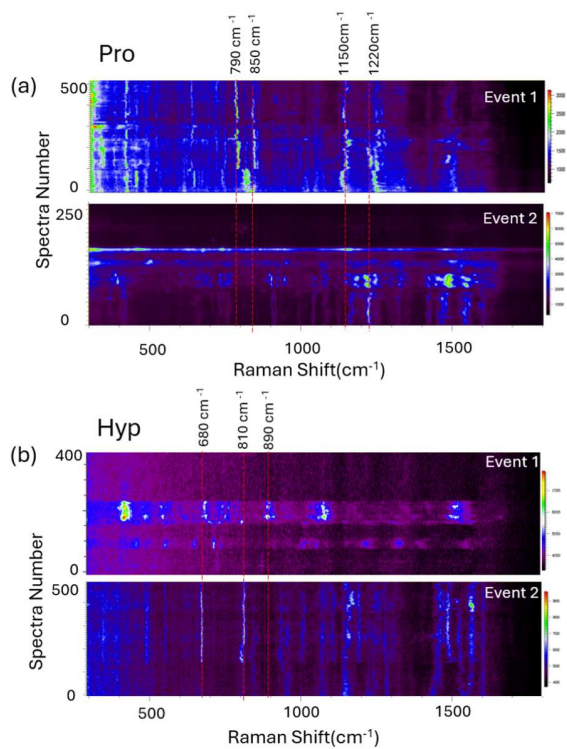

Figure 2. Waterfall plot of typical Raman spectra of (a)proline and (b)hydroxyproline trapping event. The color bars indicate the signal-to-base peak intensity.

The particle-in-pore sensor provided sensitivity at the single molecule level; however, the spectra explanation is challenging due to the spectra fluctuation and blinking, which is typical in single-molecule SERS.[22] Figure 2 presents a typical SERS data waterfall plot generated by the particle-in-pore system. Two trapping events are shown for each molecule respectively. In one trapping event, usually only part of all vibration modes can be excited. For example, in the proline event 1, three bands at $(\delta COO-)790 cm^{-1}$, $(\rho CH_2)850 cm^{-1}$, and $(\tau CH_2)1150 cm^{-1}$ can be observed but is not shown in event 2, while the band near $(\omega CH_2)1220 cm^{-1}$ is shown in both events. Similarly, for the hydroxyproline band at $(\tau Ring), 680 cm^{-1}$ can be found in both events, but the band in $(cit) 890 cm^{-1}$ and $(\nu Ring) 810 cm^{-1}$ can only be found in events 1 or 2, respectively. This is because when AuNP is trapped near the sidewall of a nanopore, the formed hot spot has the size of several angstroms. It is comparable with the size of the molecule, such that usually only part of the molecule is located in the hot spot. As a result, few vibration bands are excited depending on the relative position and the orientation of the molecule.[23] Considering the molecule would diffuse on the nanoparticle, the particle-in-pore sensor reads different moieties of the molecule at each spectrum but can survey different segments of the molecule by collecting large amounts of data and putting all puzzles together.

**Histogram to characterize spectra fluctuation.** Raman bands fluctuate in both position and intensity. The Raman peak intensity is influenced by the distance between the particle and nanopore side wall, size of hot spot, molecule position, and orientation, which provides limited information about molecule structure differences. On the other hand, even though the peak position fluctuates, the central position can be extracted statistically with sufficient spectra, thereby providing information related to structure change. By calculating the peak position and frequency of occurrence in all the obtained trapping events, the general spectral characteristics could be extracted.[24]

Peak frequency histogram was an efficient tool to obtain a general understanding of the large amount of single-molecule SERS data. [24] We counted the peak occurring frequency at each Raman shift and plotted histograms of the normalized frequency of proline and hydroxyproline, as shown in Figure 3 (a) and (b). The raw data was pre-processed by normalization, smoothing and baseline deduction. The number of peak occurrences was divided by the total number of effective spectra. (The details are listed in the experiment section) In the histograms, the general profile of proline and hydroxyproline resembles each other due to similar structure while different in detailed peak positions. Showing the distribution of the peak occurrence frequency and position, the histogram peaks reflect the position of the Raman vibration bands. With sharp bandwidth, the histogram reveals the Raman bands otherwise overlapping in spectra of solids or multi-molecule SERS. For example, in the histograms of proline, four $CH_2$ rock bands near $825 cm^{-1}$, $844 cm^{-1}$, $855 cm^{-1}$, and $878 cm^{-1}$ can be observed,[25] which may merge into two or three bands in solid or multi-molecule SERS spectra. In the particle-in-pore sensor, though not all the vibration modes are excited in one trapping event, the excited modes reveal rich and detailed spectra information that otherwise may be merged in the multi-molecule SERS spectra. With the help of the histogram of peak occurrence frequency, the citrate substitution was investigated and to determine the valid data set to be provided to the CNN model training, which will be discussed later.

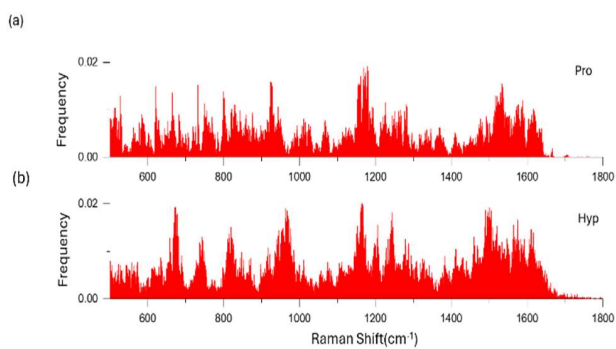

*Figure 3. Distribution histograms of Normalized peak occurrence frequency of (a)proline and (b)hydroxyproline. The number of spectra events contained in the histograms for proline and hydroxyproline 11002 and 9769 respectively.*

**Citrate substitution.** The potential interference of surfactant on AuNP makes the precise peak assignment based on the single-molecule data obtained in the particle-in-pore system very challenging. With extremely high sensitivity, the existence of citrate, which is used as a surfactant to stabilize the AuNP in our experiment, is no longer negligible. We investigated the analyte substituting on AuNPs by flowing AuNP that was incubated in the analyte solution for 0h and 48h through nanopores and calculated the peak-occurring frequency. Following our previous protocol, 48h incubation was the typical incubation time, balancing sufficient substitution and colloidal stability.[17] The analyte concentration in the incubation solution resulted in monolayer analyte coverage on AuNP. Another flow-through test using AuNP with 1/8 monolayer Hyp coverage and 24h incubation time was also performed as an example of insufficient citrate substitution. With no added analyte, the AuNPs were capped with citrate, the peak occurring frequency of which is shown in Figure 4 (a) in the black curve. Four high-intensity bands at 610 cm$^{-1}$($\delta$COO)[26], 714 cm$^{-1}$ ($\delta$OCO)[27], 1070 cm$^{-1}$($\nu$CO) [26] and 1172 cm$^{-1}$($\delta$COO)[26] were chosen to indicate the existence of citrate. The height of the four selected bands is plotted in Figure 4. (b). We observed that the calculated peak occurring frequency height is qualitatively related to the molecule population adsorbed on the surface of AuNP. In the case of sufficient substitution, the peak at 1070 cm$^{-1}$ remains obvious, while in the case of the 48h incubation monolayer, this peak height significantly decreased. In the case of insufficient substitution, the 610cm$^{-1}$ and 714cm$^{-1}$ bands' height is lower than AuNP with 0h incubation but higher than the 48h incubation. Due to the overlapping with the peak of the proline/hydroxyproline vibration mode, the 1172cm$^{-1}$ band didn't show a clear change trend. For 48h, the citrate bands became almost neglectable, indicating a minimized interference. Admittedly, this is not proof that the analyte molecule had substituted all the citrate; instead, as the analyte physically adsorbs to the AuNP in a dynamic process, even the amount of analyte in the incubating solution was enough to form a monolayer, not all the citrate can be substituted. What's more, citrate may form a second capping layer of the AuNP by hydrogen bonding [28, 29], therefore making the elimination of the citrate interference almost impossible. However, with the assistance of a deep learning model, the difference between the Pro and Hyp data can be extracted, while the residual citrate signal can be regarded as a common feature. With minimized citrate interference, the data collected from AuNP with monolayer capping and 48h incubation time was provided to the CNN model for initial training.

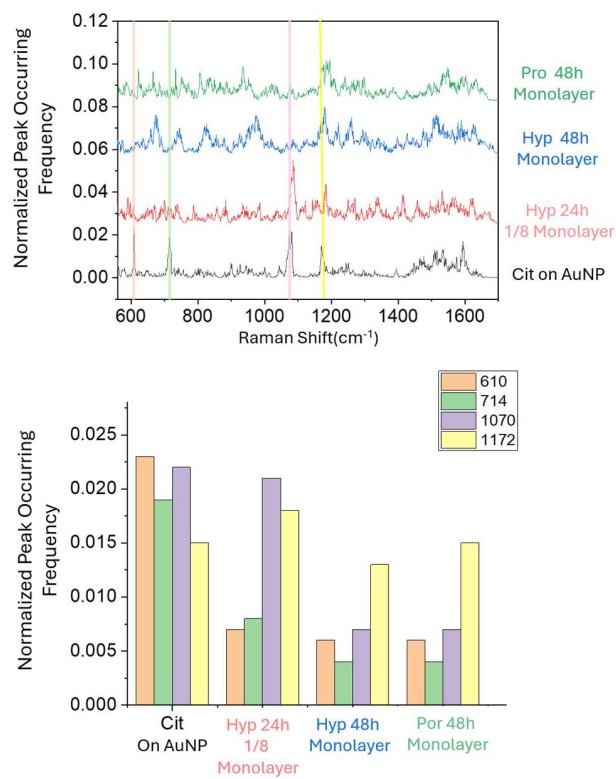

*Figure 4. (a)The normalized peak occurring frequency generated in particle-in-pore sensors using AuNP with citrate capping(black curve), 24h incubation in solution with Hyp concentration for 1/8 monolayer(red curve), 48h incubation in solution with Hyp concentration for a monolayer(blue curve), 48h incubation in solution with Pro concentration for a monolayer. Four characteristic peaks of citrate-capped*

*AuNP at 610cm$^{-1}$, 714cm$^{-1}$, 1070cm$^{-1}$ and 1172cm$^{-1}$ are indicated by orange, light green, purple and yellow guiding lines. (b) Histograms of the band height at 610cm$^{-1}$, 714cm$^{-1}$, 1070cm$^{-1}$ and 1172cm$^{-1}$, generated by AuNP with citrate capping, 24h incubation in solution with Hyp concentration for 1/8 monolayer, 48h incubation in solution with Hyp concentration for a monolayer, 48h incubation in solution with Pro concentration for a monolayer.*

**Deep learning analysis**. Deep learning-based models have been widely used in spectroscopy data analysis due to their self-adaption characterisation and high accuracy. A one-dimensional convolutional neural network model examined the prediction of proline and hydroxyproline.[30, 31] The model demonstrated accuracy higher than 96% both in the training set and post-evaluation, as shown in Figure 5. (a). It also showed tolerance to citrate interference by discriminating the data obtained from different incubation conditions. For the 72h incubation and 1/8 monolayer coverage data set, the post-evaluation accuracy is 77.9% and 85.3% for hydroxyproline and proline, respectively. (Figure 5 (c) and (d)) The lowering of the accuracy compared with the monolayer Pro or Hyp dataset might be due to interference of a larger amount of citrate remaining on AuNP. However, the 24h 1/8 monolayer Pro and Hyp dataset showed an accuracy of less than 70% (shown in Figure 5 (c) and (d)), probable because the portion of citrate signals the dominant in the dataset, which was also indicated by the peak occurrence frequency red curve in Figure 4 (a) and histogram in Figure 4 (b).

1D gradient-weighted feature visualization was an emerging tool that has been frequently used in recent works to extract spectrum features from the CNN model. [32, 33] However, many previous works used this method to analyze multi-molecule data with non-obvious spectral changes, instead of single-molecule data, which is inherently inhomogeneous. [32, 33] We made the temptation to apply it to single molecules spectra analysis and compare the extracted feature map with the peak occurring frequency. Figure 5 (b) shows the comparison of feature weight extracted from the CNN model and normalized peak occurring frequency in the region 600-1500cm$^{-1}$. The feature weight shows narrow spikes indicating high weight for the discrimination. We observe the trend that high feature weight spikes(>0.4) fall in the below regions: 750-800cm$^{-1}$, 800-825cm$^{-1}$, 900-1000cm$^{-1}$, and 1150-1200 cm$^{-1}$, which corresponds to the band difference induced by the structure change. In Table 1, the band Raman shift and vibration mode of proline and hydroxyproline are listed. In the region 750-800cm$^{-1}$, the hydroxyproline ring vibration band is at 740cm-1, while the proline shows a ring vibration-related band at 750cm$^{-1}$.[14, 15] [26] In the region 800-825 cm$^{-1}$, hydroxyproline and proline show a peak at 817cm$^{-1}$ and 824 cm$^{-1}$, which corresponds to the ring fragment CH$_2$ movement[14, 15]; the shift from 817 to 824 is caused by the structure change induced by the OH group on the ring. Similarly, at region 900-1000cm$^{-1}$, hydroxyproline and proline show peaks at 943cm$^{-1}$ and 955cm$^{-1}$, respectively, corresponding to the υRing mode.[14, 15] In region 1150-1200 cm$^{-1}$, hydroxyproline shows peaks at 1165 cm$^{-1}$ corresponding to δCH NH OH [14] and proline shows 1159 cm$^{-1}$, 1176 cm$^{-1}$ corresponding to tCH$_2$[25]Interestingly, the 1D-CNN model also gives high feature weight at positions with no peak occurrence, for example, the valley at 1036cm$^{-1}$ for proline and 1044cm$^{-1}$ for hydroxyproline. The feature weight follows the Raman band distribution in the peak-occurring frequency, indicating that the major contributions to spectral changes originate from the structure variation.

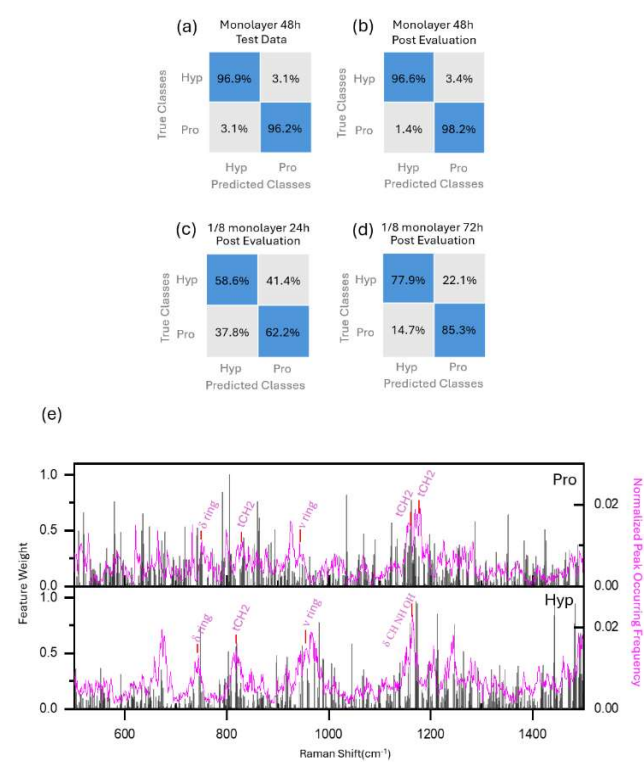

*Figure 5. Confusion matrix of the prediction accuracy for (a)training data of monolayer Hyp and Pro 48h incubation, (b)post-evaluation of monolayer Hyp and Pro 48h incubation, (c) post-evaluation of 1/8 monolayer Hyp and Pro 24h incubation and (d) post-evaluation of 1/8 monolayer Hyp and Pro 72h incubation. (e)The feature weight extracted from the 1-CNN model. The magenta curve shows the normalized peak occurring frequency.*

*Table1. The Raman bands have a high feature weight in Figure 5(e) and the corresponding vibration mode.*

| Pro Band (cm-1) | Vibration Mode |
|---|---|
| 750 | δ ring or skel deformation |
| 824 | t CH2 |
| 943 | ν ring |
| 1157,1174 | t CH2 |
| Hyp Band (cm$^{-1}$) | Vibration Mode |
| 740 | δ ring |
| 817 | t CH2 |
| 955 | ν ring |
| 1163 | δ CH NH OH |

## Conclusions

We for the first time demonstrated the discrimination of proline and hydroxylated proline at the single molecule level in the particle-in-pore sensor. The single-molecule data set's general characteristics were obtained by calculating the peak occurring frequency and plotting the frequency distribution diagrams. With the assistance of the peak occurring frequency diagram, the substitution of the citrate by analyte on AuNP was investigated, revealing that sufficient incubation in analyte solution will minimize the citrate bands in the peak frequency diagrams, which indicates the substitution of analyte on the AuNP. With the assistance of a 1D CNN model, more than 96% accuracy in discriminating proline and hydroxylated proline could be achieved. The discrimination accuracy was above 77% for 1/8 monolayer Pro and Hyp coverage, indicating that the trained 1D CNN model has a certain tolerance to citrate interference. However, with excessive citrate on the AuNP surface, in the case of the 1/8 monolayer 24h incubation dataset, the 1D CNN model can not correctly discriminate Pro from Hyp. 1D gradient-weighted feature visualization was used to output normalized feature gradients and compare them with peak occurrence frequencies. High feature weight is given to the band regions corresponding to vibration modes related to ring vibration and OH bending. The successful discrimination of proline and hydroxylated proline, which is one of the most challenging PTM detection at the single molecule level, demonstrates the great potential of our technology in various PTM discrimination. The single-molecule level sensitivity for discrimination, such as structural change with small Raman cross-session, also paves the way for site-specific PTM analysis in peptides and proteins.

## ASSOCIATED CONTENT

**Supporting Information**

The Supporting Information is available free of charge on the ACS Publications website.

## AUTHOR INFORMATION

**Corresponding Author**

Jianan Huang, Email: jianan.huang@oulu.fi

**Author Contributions**

Yingqi Zhao fabricated the Particle-in-Pore devices, determined the Raman measurement protocol, collected Raman spectra, analyzed the Raman spectra, and wrote the manuscript. Kuo wrote the scripts for Raman data pre-processing, effective spectrum selecting, and peak occurrence frequency calculation and developed a 1D-CNN model for PTM discrimination. Pei-Lin Xin contributed to writing the Raman spectra pre-processing and data selection scripts. Francesco De Angelis contributed to the device fabrication and revised the manuscript. Jianan Huang conceived the idea, acquired funding, supervised the work, and revised the manuscript. The manuscript was written with contributions from all authors. All authors have approved the final version of the manuscript.

## ACKNOWLEDGMENT

This research receives support from Academy Research Fellow projects: TwoPoreProSeq (project number 347652), Biocenter Oulu emerging project (DigiRaman) and DigiHealth project (project number 326291), a strategic profiling project at the University of Oulu that is supported by the Academy of Finland and the University of Oulu.